# Quantum attomicroscopy:
# imaging quantum chemistry in action

Nikolay V. Golubev* and Mohammed Th. Hassan*

Department of Physics, University of Arizona, Tucson, AZ 85721, USA.

*Corresponding authors: ngolubev@arizona.edu and mohammedhassan@arizona.edu

**Abstract**

**How quantum electron and nuclei motions affect biomolecular chemical reactions remains a central challengeable question at the interface of quantum chemistry and biology. Ultrafast charge migration in deoxyribonucleic acid (DNA) has long been hypothesized to play a critical role in photochemistry, genome stability, and long-range biomolecular signaling; however, direct real-time observation of these electronic processes has remained elusive. Here, we present a theoretical investigation and propose the concept of future experimental measurements of laser-driven charge dynamics in the canonical DNA nucleobase pairs: thymine–adenine and cytosine–guanine. Attosecond-resolved simulations employing high-level *ab initio* methods reveal base-dependent ionization mechanisms, directional charge-migration pathways, and electronic coherences that govern sub-femtosecond redistribution of electron density across hydrogen-bonded nucleobase interfaces. Accordingly, we propose the concept of a quantum attosecond scanning electron microscope, termed the quantum attomicroscope (Q-attomicroscope), a capable of imaging photoinduced quantum chemistry reactions in attosecond temporal resolution and sub-nanometer spatial precision. As a proof of principle, we propose to image the charge migrations dynamics in DNA which we studied theoretically. Together, our preceptive bridges theory, instrumentation, and control, outlining a pathway toward laser-mediated manipulation of DNA structure with implications for repair processes, chemical reactivity, and future personalized medicine.**

# 1- Introduction

For much of the twentieth century, chemical processes were assumed to occur on timescales far too short to be directly observed [1]. This perception changed dramatically with the advent of femtosecond time-resolved spectroscopy. By effectively "freezing time," this technique enabled experimental access to the structural dynamics of chemical reactions, providing unprecedented insight into transition states, coherent control, and reaction outcomes [2]. These developments gave rise to the field of ultrafast science and femtochemistry, in which atomic and molecular motion can be monitored in real time [3].

At their most fundamental level, chemical reactions are governed by the breaking and formation of chemical bonds. These processes are driven by electron motion, which naturally unfolds on femtosecond to attosecond timescales. As a result, the direct observation—and ultimately the control—of chemical reactions as they occur has long remained beyond experimental reach.

Over the past decade, major advances in attosecond science have begun to overcome this limitation. In particular, the generation of attosecond extreme ultraviolet (XUV) pulses via high-harmonic generation has enabled real-time observation of electron dynamics in atoms, molecules, and solids [4-8]. More recently, attosecond time-resolved spectroscopy has been used to probe charge migration in molecules [6,9,10], inaugurating the emerging field of attochemistry [11-17]. Despite these breakthroughs, most existing studies rely on high-energy XUV photons and primarily probe charge-transfer dynamics in dissociating molecular fragments rather than in intact molecular systems. Furthermore, while attosecond spectroscopy provides valuable temporal information on transient electronic states, it offers limited insight into the spatial pathways and mechanisms of electron motion. These limitations underscore the need for new experimental tools capable of directly imaging chemical reactions and electron dynamics in space and time.

In this perspective, we propose such a tool, which we term the quantum attomicroscope (Q-attomicroscope). The concept is based on extending scanning tunnelling microscopy by inducing the tunnelling current with a half-cycle laser pulse confined to the sub-femtosecond regime. This approach aims to achieve simultaneous attosecond temporal and angstrom-scale spatial resolution, thereby enabling direct imaging of electron motion during chemical reactions. The underlying principles and proposed implementation of the Q-attomicroscope are discussed in detail in Section 3. The ability to image electron dynamics in real space and time would open transformative opportunities across chemistry, biology, and materials science. In particular, visualizing charge migration in molecules and biomolecules could provide critical insight into fundamental processes such as charge transfer and electron localization. For example, imaging charge migration in amino acids—the building blocks of proteins—would inform our understanding of electron dynamics in more complex biological systems, including deoxyribonucleic acid (DNA) and proteins. Such capabilities could address longstanding questions in molecular biology, such as whether electrons tunnel between the two strands of DNA, along which trajectories, and on what timescales. Resolving these questions would deepen our understanding of molecular interactions, including DNA–protein interactions during transcription and translation processes, with direct implications for carcinogenesis, mutagenesis, and DNA repair mechanisms [18]. Beyond biology, these insights could also guide the development of DNA-based molecular electronics and bioinformatics applications [19].

More broadly, the principle that control follows observation suggests that Q-attomicroscopy could ultimately enable the manipulation of photochemical reactions and molecular conformations in real space. Such capabilities would represent a paradigm shift, with potential long-term implications for personalized medicine and targeted molecular control.

As an initial step toward realizing this vision, we present a theoretical investigation of charge migration in the two fundamental DNA nucleobase pairs: thymine–adenine (T–A) and cytosine–guanine (C–G). We discuss the working principle and potential usage of Q-attomicroscope for observing the ultrafast electron dynamics in these systems. The presented combined theoretical and computational studies provide guidance on optimal experimental parameters and anticipated observables, laying the groundwork for future experiments using the Q-attomicroscope.

## 2- Theoretical study of the charge migration dynamics in DNA nucleobase pairs

Theoretical treatment of ultrafast quantum dynamics in complex molecular systems such as DNA base pairs is highly challenging. A central difficulty in performing accurate numerical simulations lies in the need to account for the strongly correlated motion of a large number of electrons and nuclei forming the pairs. While a fully non-adiabatic description of the coupled electron-nuclear dynamics remains far beyond current computational capabilities for systems of this size, the electronic structure and the ultrafast electron motion can often be treated within well-established quantum-chemical frameworks by assuming fixed or parametrized nuclear configurations. In fact, unprecedented temporal resolution of modern experimental techniques such as Q-attomicroscope proposed in this perspective can be sufficient to capture the pure electron motion in the system before the nuclear rearrangement comes into play. Accordingly, in the follow-up discussion we present fully *ab initio* numerical simulations of the electronic structure and dynamics in T–A and C–G base pairs studying the motion of electron density triggered by the interaction of these systems with intense ultrashort laser pulses.

Our calculations consisted of the following steps. First, preliminary molecular geometries of the nucleobases were constructed using the simplified molecular-input line entry system (SMILES) format using OpenBabel 3.0.0[20] with the MMFF94 force field[21]. The ground-state geometries of T, A, C, and G neutral molecules were optimized using the density functional theory (DFT) at the B3LYP[22]/aug-cc-pVDZ[23] level. The obtained individual molecules were arranged in the corresponding T–A and C–G pairs and the entire structures were re-optimized at the same level of theory.

In the next step, we computed the ionic spectra. The noncorrelated reference orbitals were obtained using the Hartree–Fock (HF) method with aug-cc-pVDZ basis set. The cationic Hamiltonian was constructed using the non-Dyson version of the algebraic diagrammatic construction (ADC) scheme[24,25] at the third order of perturbation theory for representing the one-particle Green's function. The active space was chosen to contain all the orbitals except $1s$ core states of all heavy atoms. All one-hole (1h) and two-hole-one-particle (2h1p) electronic configurations were taken into account in the calculations of ionic states and in the follow-up time-dependent density analysis. All simulations have been performed with the Q-Chem software package[26].

Ionizing an electron from T–A and C–G base pairs, we were searching for conditions under which the correlation between the electrons of neutral systems causes the so-called hole-mixing effect[27,28]. This effect arises when certain electrons of a molecule are particularly strongly correlated with each other such that localized ionization of a single electron causes the rearrangement of other electrons in the system. Figs. 1a and 2a show the calculated ionization spectra of T–A and C–G base pairs, respectively. Vertical bars represent ionic states with different colors depicting the localization of the created hole in the molecular orbitals of the neutral system. It is seen that most of the low-lying ionic states for both systems are composed of a single configuration indicating that ionization of an electron from outer-valence orbitals will not launch any electron dynamics. However, energetically deeper ionic states demonstrate a strong evidence of the hole-mixing between the orbitals of the neutral system. Accordingly, ionizing an electron from one of these inner-valence orbitals, one will unavoidably involve the other electron from the corresponding orbital in the mixture thus launching the ultrafast dynamics of the electron density [29,30]. The time scale of these dynamics is determined by the energy separation between the electronic states participating in the hole mixing. In case of T–A base pair, the energy separation between the states is found to be ~0.4 eV which corresponds to ~10.5 fs oscillation period. In C–G, the states are separated by 0.165 eV which leads to slower dynamics with ~25 fs period of oscillations.

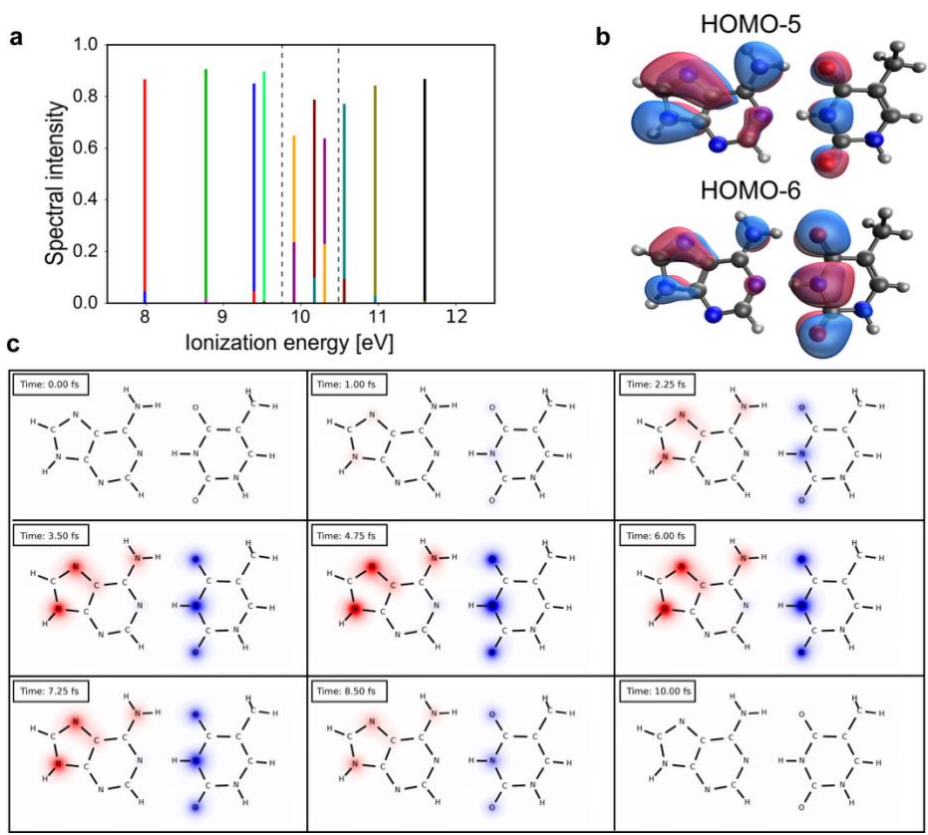

**Figure 1. Charge migration dynamics in thymine–adenine nucleobase pair**. **a**, Ionization spectrum of thymine–adenine nucleobase pair. It is seen that the inner ionic states of the system experience strong hole-mixing effect between the molecular orbitals HOMO–5 and HOMO–6 depicted in **b**. **c**, The evolution of the electron difference density triggered by the sudden removal of an electron from HOMO–5 orbital of the system. It is seen that the localized ionization launches the electronic oscillations between molecules in the pair.

Investigating the nature of the orbitals involved in hole mixing in T–A and C–G base pairs, we were surprised to find that in both cases these orbitals are delocalized on the molecules forming the pairs, as shown in Figs. 1b and 2b. Therefore, the electron dynamics triggered by ionizing an appropriate inner electron will show up in the form of the oscillations of the electron density between the individual molecules in the corresponding pair. To the best of our knowledge, the existence of the hole-mixing and the associated charge migration dynamics in non-covalently bound parts of the system has not been reported to date.

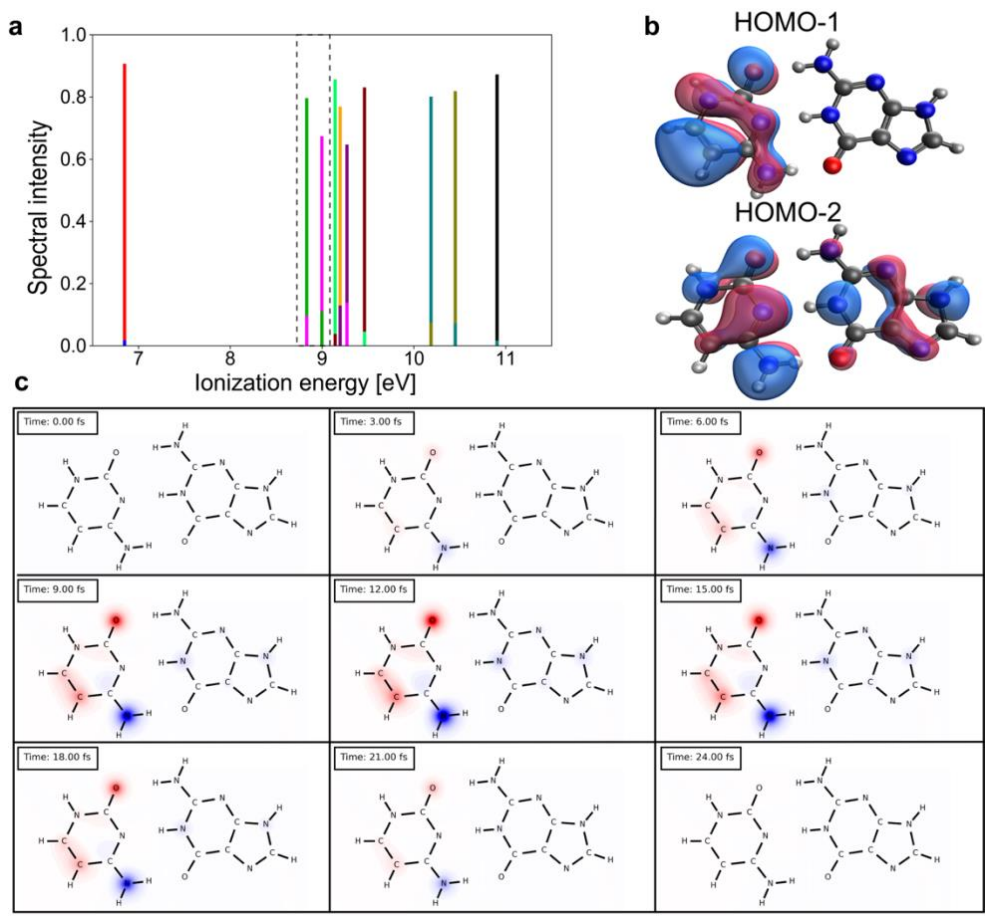

**Figure 2. Charge migration dynamics in cytosine–guanine nucleobase pair.** a, Ionization spectrum of cytosine–guanine nucleobase pair. Similarly to the case of thymine–adenine depicted in Fig. 1, the ionic states of the system experience strong hole-mixing effect between the molecular orbitals HOMO–1 and HOMO–2 depicted in b. c, The evolution of the electron difference density triggered by the sudden removal of an electron from HOMO–1 orbital of the system. In contrast to Fig. 1, the initiated electron dynamics primarily takes place in a single molecule in the pair.

We simulated the dynamics of the electron density in T–A and C–G base pairs following the sudden removal of an electron from HOMO–5 and HOMO–1 orbitals, respectively. Figs. 1c and 2c demonstrate the evolution of the electron difference density:

$$\Delta Q(\mathbf{r}, t) = Q(\mathbf{r}, t) - Q(\mathbf{r}, 0),$$

where $\mathbf{r}$ denotes the coordinate of an electron and $t$ is time variable. The corresponding hole density $Q(\mathbf{r}, t)$ has been computed as

$$Q(\mathbf{r}, t) = \sum_{p,q} \varphi_p^*(\mathbf{r}) \varphi_q(\mathbf{r}) N_{pq}(t),$$

where $\varphi_p(\mathbf{r})$ and $\varphi_q(\mathbf{r})$ are the molecular orbitals of the neutral system, and $N_{pq}(t)$ is the so-called hole density matrix[27,28]. Accordingly, the positive values of $\Delta Q(\mathbf{r}, t)$ quantity indicate that the electron density increases over time in certain regions of space, while the negative values denote the decrease of the electron density with respect to the initial one. In case of T–A base pair, it is very well seen (see Fig. 1c) that the electronic oscillations take place between T and A molecules. In case of C–G pair, shown in Fig. 2c, the contrast of intra-molecular oscillations is found to be rather weak, which originates from weaker hole-mixing between the corresponding molecular orbitals (see Fig. 2a).

Our theoretical predictions provide us with strong evidence that the electron correlation in nucleobase pairs can lead to ultrafast intra-molecular dynamics of the electron density between the molecules in the pairs. However, the complexity of the nucleobase pairs, and DNA and biomolecules in general, prohibits the complete fully quantum analysis and simulations of such systems. To overcome this obstacle, we describe in the next section the perspective to utilize Q-attomicroscope technique which can help to unravel the quantum dynamics in large biologically relevant systems.

3- **The development of quantum attosecond electron microscope (Q-Attomicroscope)**

Earlier, we developed attosecond electron microscopy, which we coined the attomicroscope[31], based on the transformation of a transmission electron microscope (TEM). This new tool enabled attomicroscopy electron imaging of graphene via electron diffraction. However, capturing charge migration at the subatomic level, such as in DNA base pairs (Figs. 1 and 2), requires the simultaneous combination of attosecond temporal and angstrom spatial resolution. To this end, we are developing the quantum version of the attomicroscope (Q-attomicroscope), aimed at achieving attosecond temporal resolution within a scanning tunnelling microscope (STM).

The Q-attomicroscope (illustrated in Fig. 3a) is based on generating a light-induced attosecond tunnelling current in a modified STM. This can be achieved either by illuminating the STM tip with optical attosecond pulses (OAPs)[32-34] or by using a polarization-gated half-cycle pulses (PGPs), similar to the approach employed in the development of Attomicroscope 1.0[31]. The PGP is generated by sending a 1.5-cycle laser pulse through a combination of zero- and multiple-order waveplates, producing circular polarization at the pulse edges and a linearly polarized half-cycle portion at the pulse centre (illustrated in Fig. 3b). On the other side, the OAP approach enables the generation of a half-cycle pulse with temporal resolution down to approximately 400 as, based on light-field synthesis [32,33]. Achieving optimal contrast between the central half-cycle and adjacent field crests typically requires several optimization iterations to ensure that tunnelling is strictly confined to the 400 as window and that no residual current is generated from neighbouring half-cycles. For this reason, the use of few-cycle pulses alone does not yield a truly isolated attosecond current signal, but rather a train of attosecond probe current events, limiting realistic attosecond imaging. In contrast, the PGP approach is more promising, as it originates from a single laser source rather than multiple spectral channels, as required in OAP synthesis. It exploits the fact that tunnelling occurs only during the linearly polarized half-cycle portion of the pulse, thereby enabling sub-femtosecond-resolution tunnelling current.

It is worth noting that optically induced tunnelling currents suffer from substantial signal noise, since laser pulse amplitude noise is multiplicatively amplified in the tunnelling process. However, recent developments in ultrafast quantum optics within Hassan group [35,36] enabled the use of amplitude-squeezed light, which promises a significant reduction in light-induced current noise [35]. The Q-attomicroscope employs a time-resolved stroboscopic approach to effectively freeze time and capture attosecond electron motion in action. In this scheme, electron dynamics are initiated by a separate pump laser pulse focused on the sample to excite the targeted processes. During Q-attomicroscopy imaging experiments, images are recorded as a function of the time delay between the tunnelling current probe and the pump laser pulse, controlled via a delay stage in the pump beam path. By systematically varying this delay, a sequence of snapshots is obtained and subsequently stacked to reconstruct attosecond electron movies.

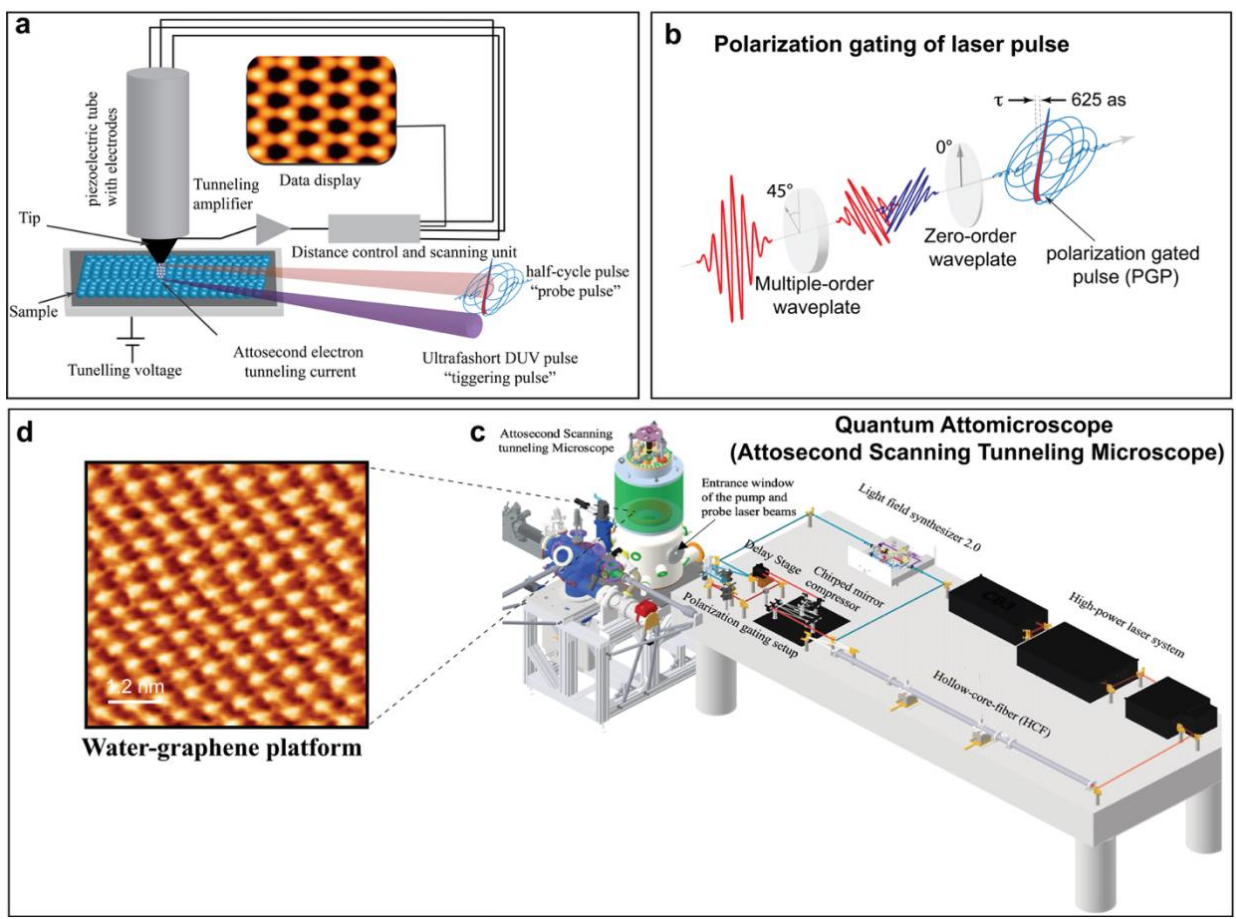

**Figure 3. Quantum attomicroscope (Q-attomicroscope). a**, Schematic illustration of the operating principle of the Q-attomicroscope, which extends a scanning tunnelling microscope (STM) to the attosecond regime. A half-cycle laser pulse is used to generate an attosecond electron tunnelling current, enabling the imaging of photoinduced electronic and chemical dynamics initiated by a synchronized ultrashort deep-ultraviolet (DUV) pump pulse. Time-resolved STM images are obtained by scanning the sample with controlled pump–probe delay. **b**, Principle of generating a half-cycle, polarization-gated laser pulse using ultrafast laser techniques, which is employed to drive attosecond tunnelling currents in the STM junction. **c**, Complete Q-attomicroscope experimental setup, from the laser system to the STM end station, designed to image charge migration dynamics in molecular systems. **d**, Experimental STM image of a graphene/water platform, which will serve as a native-like environment for future charge-migration imaging of DNA base pair molecules.

Next, we describe the use of polarization-gated pulses (PGP) to develop the Q-attomicroscope for imaging charge migration in DNA base pairs (T–A and C–G), which were theoretically studied and presented earlier in this article. The Q-attomicroscope experimental setup is shown in Fig. 3c. The system is driven by a high-power, high-repetition-rate (100 kHz) laser combined with optical

- 9 -

parametric chirped-pulse amplification (OPCPA) to generate 10 fs pulses with an average power of 10 W. The output beam is focused into a neon-filled hollow-core fiber (HCF) to generate a broadband supercontinuum spanning 250–1000 nm. This supercontinuum is subsequently divided into two beams. The first beam is compressed using a chirped-mirror compressor to 5 fs. This beam is then reflected from a rectangular reflector mounted on a precision delay stage (Fig. 3c). This delay stage has a nanometre resolution and provides a delay steps in the order of ten attoseconds. This beam will be sent through a polarization-gating setup to generate a linearly polarized half-cycle gating pulse. The resulting PGP is focused onto the STM tip to generate a light-induced attosecond tunnelling current, which serves as the probe for imaging and recording the charge-migration dynamics.

The second beam is directed to the second generation of the Attosecond Light-Field Synthesizer (ALFS 2.0), which is being developed to generate a short, intense deep-ultraviolet (DUV) pump pulse. ALFS 2.0 will incorporate two DUV spectral channels. The first channel (DUV1) spans 500–350 nm, similar to the channel demonstrated in ALFS 1.0 [33,37] and can deliver up to 200 mW with a pulse duration of 5 fs. The second channel spans 350–250 nm, providing lower power but a comparable pulse duration. The two DUV channels will be coherently combined to form a 5 μJ, 3 fs DUV pulse with a central photon energy of 3.54 eV. This pulse will be tightly focused to a 5 μm spot size, yielding a peak intensity on the order of a few to tens of $10^{13}$ W/cm$^2$. Such intensity is sufficient to initiate charge-migration dynamics in DNA base pairs via three-photon absorption, reaching excitation energies of approximately 10 eV (see Figs. 1b and 2b). Direct removal of an electron from isolated DNA base pairs (T–A or C–G) would lead to irreversible sample damage. To mitigate this, the DNA bases will be supported on a monolayer of water molecules, providing a stabilizing environment. This $H_2O$ monolayer can be realized by cooling a graphene substrate to 80 K, enabling the formation of an ultrathin water sheet. This approach provides a realistic aqueous-like environment, allowing charge migration in DNA base pairs to be studied under near-native conditions. The ultimate goal of this proposed experimental study is to record experimentally the attosecond charge migration movies similar to the simulated dynamics presented in animations as supplementary material videos 1 and 2.

It is worth noting that the development of the Q-attomicroscope presents several technical challenges. Among these, the most critical is the noise associated with the photoinduced tunnelling current, which adds to the intrinsic dark noise of the STM, such as thermal noise. We aim to overcome this limitation by employing half-cycle, amplitude-squeezed light, as we have recently demonstrated[36]. In quantum-light–induced tunnelling, this approach enables a substantial reduction of current noise, enhancing the signal-to-noise ratio of the generated tunnelling current by nearly an order of magnitude [35].

## 4- Summary and outlook

In summary, we propose the development of a Q-attomicroscope by adopting a time-resolved scanning tunnelling microscopy (STM) approach combined with half-cycle laser pulses to generate sub-femtosecond tunnelling current signals. This capability would provide simultaneous attosecond temporal and angstrom spatial resolution, thereby opening the door to directly imaging electron motion and the initiation of chemical reactions in real time. Beyond observation, we envision that this capability will enable the use of tailored light fields [32,33,38] to actively control chemical reactions—a long-anticipated goal in the scientific community. Such control over chemistry would have far-reaching implications for applications in biology and materials science. As a proof of principle for the Q-attomicroscope, we propose to image attosecond-resolved movies of charge migration in DNA base-pair molecules. To support this effort, we have performed theoretical calculations to identify the optimal experimental parameters (e.g., photon energy) and to predict the expected electron dynamics. Together, these results establish a clear pathway toward realizing attosecond-resolved, real-space imaging and control of fundamental chemical processes.

**Acknowledgments**

NG acknowledges the financial support from the U.S. Department of Energy (DOE), Office of Science, Basic Energy Sciences (BES) under Award #DE-SC0024182. MH would like to thank the Air Force Office of Scientific Research under award number FA9550-22-1-0494 for supporting the development of Q-attomicroscope and a Gordon and Betty Moore Foundation Grant for funding the Imaging Electron Motion and Electron Driven Chemical Reactions at Single Atom Resolution project (GBMF 11476, https://www.moore.org/grant-detail?grantId=GBMF11476 ).

# References

1. Eigen, M. Immeasurably fast reactions. *Nobel Lecture* **11**, 1963-1979, (1967).
2. Zewail, A. H. Femtochemistry: Atomic-scale dynamics of the chemical bond. *The Journal of Physical Chemistry A* **104**, 5660-5694, (2000).
3. Zewail, A. H. Femtochemistry: Atomic-scale dynamics of the chemical bond using ultrafast lasers (Nobel Lecture). *Angewandte Chemie International Edition* **39**, 2586-2631, (2000).
4. Corkum, P. & Krausz, F. Attosecond science. *Nat. Phys.* **3**, 381-387, (2007).
5. Krausz, F. & Ivanov, M. Attosecond physics. *Rev. Mod. Phys.* **81**, 163, (2009).
6. Calegari, F. *et al.* Ultrafast electron dynamics in phenylalanine initiated by attosecond pulses. *Science* **346**, 336-339, (2014).
7. Sansone, G. *et al.* Electron localization following attosecond molecular photoionization. *Nature* **465**, 763-766, (2010).
8. Chakraborty, N. *et al.* Attosecond Transient Absorption Study of Coherent Hole Oscillation in Ar+. *arXiv preprint arXiv:2508.10261*, (2025).
9. Belshaw, L. *et al.* in *2013 Conference on Lasers & Electro-Optics Europe & International Quantum Electronics Conference CLEO EUROPE/IQEC.* 1-1.
10. Matselyukh, D. T., Despré, V., Golubev, N. V., Kuleff, A. I. & Wörner, H. J. Decoherence and revival in attosecond charge migration driven by non-adiabatic dynamics. *Nat. Phys.* **18**, 1206-1213, (2022).
11. Calegari, F. & Martin, F. Open questions in attochemistry. *Communications Chemistry* **6**, 184, (2023).
12. Nisoli, M., Decleva, P., Calegari, F., Palacios, A. & Martín, F. Attosecond Electron Dynamics in Molecules. *Chemical Reviews* **117**, 10760-10825, (2017).
13. Merritt, I. C. D., Jacquemin, D. & Vacher, M. Attochemistry: Is Controlling Electrons the Future of Photochemistry? *The Journal of Physical Chemistry Letters* **12**, 8404-8415, (2021).
14. Tran, T., Ferté, A. & Vacher, M. Simulating Attochemistry: Which Dynamics Method to Use? *The Journal of Physical Chemistry Letters* **15**, 3646-3652, (2024).
15. Folorunso, A. S. *et al.* Attochemistry Regulation of Charge Migration. *The Journal of Physical Chemistry A* **127**, 1894-1900, (2023).
16. Palacios, A. & Martín, F. The quantum chemistry of attosecond molecular science. *WIREs Computational Molecular Science* **10**, e1430, (2020).
17. Cruz-Rodriguez, L., Dey, D., Freibert, A. & Stammer, P. Quantum phenomena in attosecond science. *Nature Reviews Physics* **6**, 691-704, (2024).
18. Holmlin, R. E., Dandliker, P. J. & Barton, J. K. Charge transfer through the DNA base stack. *Angewandte Chemie International Edition in English* **36**, 2714-2730, (1997).
19. Giese, B., Amaudrut, J., Köhler, A.-K., Spormann, M. & Wessely, S. Direct observation of hole transfer through DNA by hopping between adenine bases and by tunnelling. *Nature* **412**, 318-320, (2001).
20. O'Boyle, N. M. *et al.* Open Babel: An open chemical toolbox. *Journal of cheminformatics* **3**, 33, (2011).
21. Halgren, T. A. Merck molecular force field. I. Basis, form, scope, parameterization, and performance of MMFF94. *Journal of computational chemistry* **17**, 490-519, (1996).
22. Becke, A. D. Density-functional thermochemistry. III. The role of exact exchange. *The Journal of chemical physics* **98**, 5648-5652, (1993).

23	Kendall, R. A., Dunning Jr, T. H. & Harrison, R. J. Electron affinities of the first-row atoms revisited. Systematic basis sets and wave functions. *The Journal of chemical physics* **96**, 6796-6806, (1992).
24	Schirmer, J., Cederbaum, L. S. & Walter, O. New approach to the one-particle Green's function for finite Fermi systems. *Phys. Rev. A* **28**, 1237, (1983).
25	Schirmer, J., Trofimov, A. & Stelter, G. A non-Dyson third-order approximation scheme for the electron propagator. *The Journal of chemical physics* **109**, 4734-4744, (1998).
26	Epifanovsky, E. *et al.* Software for the frontiers of quantum chemistry: An overview of developments in the Q-Chem 5 package. *The Journal of chemical physics* **155**, (2021).
27	Breidbach, J. & Cederbaum, L. Migration of holes: Numerical algorithms and implementation. *The Journal of chemical physics* **126**, (2007).
28	Breidbach, J. & Cederbaum, L. Migration of holes: Formalism, mechanisms, and illustrative applications. *The Journal of chemical physics* **118**, 3983-3996, (2003).
29	Cederbaum, L. S. & Zobeley, J. Ultrafast charge migration by electron correlation. *Chemical Physics Letters* **307**, 205-210, (1999).
30	Remacle, F., Levine, R. & Ratner, M. Charge directed reactivity:: a simple electronic model, exhibiting site selectivity, for the dissociation of ions. *Chemical physics letters* **285**, 25-33, (1998).
31	Hui, D., Alqattan, H., Sennary, M., Golubev, N. V. & Hassan, M. T. Attosecond electron microscopy and diffraction. *Science Advances* **10**, eadp5805, (2024).
32	Hassan, M. T. *et al.* Optical attosecond pulses and tracking the nonlinear response of bound electrons. *Nature* **530**, 66-70, (2016).
33	Alqattan, H., Hui, D., Pervak, V. & Hassan, M. T. Attosecond light field synthesis. *APL Photonics* **7**, 041301, (2022).
34	Alqattan, H., Hui, D., Sennary, M. & Hassan, M. T. Attosecond electronic delay response in dielectric materials. *Faraday Discuss.* **237**, 317-326, (2022).
35	Mohamed Sennary, Javier Rivera-Dean, Maciej Lewenstein & Hassan, M. T. Ultrafast Quantum Optics (2026).
36	Sennary, M. *et al.* Attosecond quantum uncertainty dynamics and ultrafast squeezed light for quantum communication. *Light: Science & Applications* **14**, 350, (2025).
37	Hui, D. *et al.* Attosecond electron motion control in dielectric. *Nat. Photon.* **16**, 33-37, (2022).
38	Hassan, M. T. Electron microscopy for attosecond science. *Physics Today* **77** 38–43 (2024).